# Upper critical field of KFe$_2$As$_2$ under pressure: A test for the change in the superconducting gap structure

Valentin Taufour,[1, 2, *] Neda Foroozani,[3] Makariy A. Tanatar,[1, 2] Jinhyuk Lim,[3] Udhara Kaluarachchi,[1] Stella K. Kim,[1, 2] Yong Liu,[2] Thomas A. Lograsso,[2] Vladimir G. Kogan,[2] Ruslan Prozorov,[1, 2] Sergey L. Bud'ko,[1, 2] James S. Schilling,[3] and Paul C. Canfield[1, 2]

[1]*Department of Physics and Astronomy, Iowa State University, Ames, Iowa 50011, U.S.A.*
[2]*The Ames Laboratory, US Department of Energy, Iowa State University, Ames, Iowa 50011, USA*
[3]*Physics Department, Washington University, St. Louis, Missouri 63130, U.S.A.*
(Dated: June 24, 2014)

We report measurements of electrical resistivity under pressure to 5.8 GPa, magnetization to 6.7 GPa, and ac susceptibility to 7.1 GPa in KFe$_2$As$_2$. The previously reported change of slope in the pressure dependence of the superconducting transition temperature $T_c(p)$ at a pressure $p^* \sim 1.8$ GPa is confirmed, and $T_c(p)$ is found to be nearly constant above $p^*$ up to 7.1 GPa. The $T$-$p$ phase diagram is very sensitive to the pressure conditions as a consequence of the anisotropic uniaxial pressure dependence of $T_c$. Across $p^*$, a change in the behavior of the upper critical field is revealed through a scaling analysis of the slope of $H_{c_2}$ with the effective mass as determined from the $A$ coefficient of the $T^2$ term of the temperature-dependent resistivity. We show that this scaling provides a quantitative test for the changes of the superconducting gap structure and suggests the development of a $k_z$ modulation of the superconducting gap above $p^*$ as a most likely explanation.

Since the discovery of superconductivity in LaFeAs(O$_{1-x}$F$_x$) [1], the iron-based superconductors have been the focus of numerous experimental and theoretical studies. Taking advantage of the lessons learned from the cuprate high-temperature superconductors, the investigation of the symmetry of the superconducting state has been given priority [2]. However, unlike the cuprates, the gap structure of the iron-based superconductors is not universal and several gap symmetries have been proposed both experimentally and theoretically [3]. The stoichiometric compound KFe$_2$As$_2$, which has a superconducting transition temperature $T_c \approx 3.5$ K, is one of the cleanest examples where different gap structures appear likely [4–15]. Recently, it has been suggested that a change of pairing symmetry from $d$ wave to $s$ wave occurs upon applying pressure to KFe$_2$As$_2$ [9]. The argument was based on the experimental observation of a change in the pressure dependence of $T_c$ from negative to positive at $p^* \approx 1.8$ GPa. Following this study, ac magnetic susceptibility and de Haas–van Alphen (dHvA) oscillations under pressure confirmed the change of slope in $T_c(p)$ at $p^*$ and supported the earlier inference that this change is not due to drastic modifications of the Fermi surface [16]. A similar change of slope of $T_c(p)$ was also observed in CsFe$_2$As$_2$ [17]. Although there have been theoretical predictions that the $d$-wave and $s$-wave states are very close in energy [14, 18, 19], the experimental data available so far do not provide information about the changes in the superconducting gap function at $p^*$.

In this study we significantly extend the pressure range of previous studies ($\sim 2.5$ GPa [9, 16]) to 7.1 GPa. We confirm the observation of a change of slope in $T_c(p)$ but find that the phase diagram is very sensitive to the hydrostaticity of the pressure medium in this high-pressure range. In addition, we report on the temperature dependence of the upper critical field $H_{c_2}$ with the magnetic field applied along the $c$ axis under pressure. By scaling $\left(-d\mu_0 H_{c_2}/dT|_{T_c}\right)/T_c$ versus the $A$ coefficient of the $T^2$ term of the resistivity, we find a change at $p^*$ which allows for a quantitative test of the modification of the superconducting gap structure. The present data are not able to test whether a change in symmetry between $d$ wave and $s$ wave actually takes place at 1.8 GPa. We find, however, that such a change alone would not be sufficient to account for our results. We suggest that a $k_z$ modulation of the superconducting gap is involved in the slope change at $p^*$.

The Fermi surface of KFe$_2$As$_2$ has been investigated experimentally by dHvA oscillation [16, 20] and angle-resolved photoemission spectroscopy (ARPES) [4, 21, 22] experiments. The Fermi surface consists of three hole cylinders at the $\Gamma$ point [$\alpha$ (inner), $\zeta$ (middle), and $\beta$ (outer) bands], and four small hole cylinders near the $X$ point ($\epsilon$ band). ARPES experiments down to 2 K indicate that the gap is nodeless on the $\alpha$ and $\beta$ bands, and nodal with octet line nodes on the $\zeta$ (middle) band [4]. The nodes have also been detected by thermal conductivity [5, 6], penetration depth [7], and nuclear quadrupole resonance [8]. The question of whether those nodes are accidental with an $s$-wave state [4] or imposed by symmetry in a $d$-wave state [6, 9] is still under debate [10]. Other possibilities include a time-reversal symmetry breaking $s + id$ state [11–14], or an $s + is$ state between two kinds of $s\pm$ states which has been proposed upon Ba doping in Ba$_{1-x}$K$_x$Fe$_2$As$_2$ [15] in the vicinity of $x \sim 0.7$ where deviations in the jump in specific heat have been observed [23, 24]. In this context, the evidence for a change of gap function under pressure in $KFe_2As_2$ illustrates the near degeneracy of these states and the possibility of studying the interplay between different superconducting states. Our results suggest that, in addition to the considered possible in-plane symmetry of the gap functions, a $k_z$ modulation of the superconducting gap is involved in the slope change at $p^*$.

In this study, several high-quality single crystals of $KFe_2As_2$ were grown from KAs flux as detailed in Ref. 25. For the electrical resistivity measurements, pressure was applied at room temperature using a modified Bridgman cell [26] with a 1:1 mixture of n-pentane:isopentane as a pressure medium. The ac susceptibility measurements to hydrostatic pressures as high as 7.1 GPa were carried out in a membrane-loaded diamond-anvil cell [27]. Helium was used as pressure medium. To promote hydrostaticity, pressure was increased at temperatures well above the melting curve of helium, unless stated otherwise. Further experimental details are given in the Supplementary Material for further experimental details and other measurements using media with less hydrostatic pressure.[28](see also Refs. [9, 16, 26, 27, 29–49]) together with other measurements using less hydrostatic pressure media.

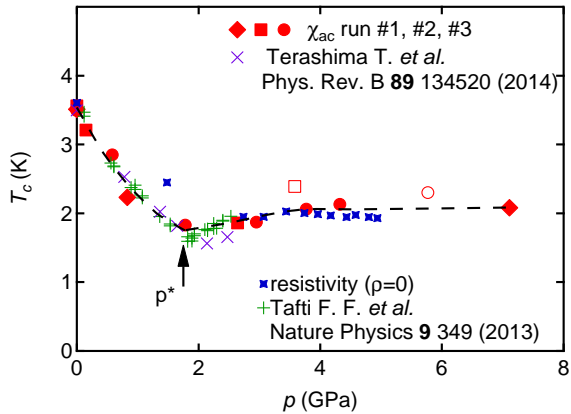

Figure 1. (Color online) Superconducting phase diagram for $KFe_2As_2$ determined from our resistivity and ac susceptibility measurements [28]. For the latter, filled symbols are used when pressure was applied on liquid helium, whereas open symbols are used when pressure was applied on solid helium (at 10 K). The dashed line is a guide to the eye which does not use the two data points where the pressure was applied on solid helium. The data from Refs. [9, 16] are also shown.

The superconducting phase diagram obtained from ac susceptibility and resistivity measurements is shown in Fig. 1. The previously reported change of slope in $T_c(p)$ at $p^* \approx 1.8$ GPa [9, 16] is confirmed. $T_c$ increases very slowly above $p^*$ up to 7.1 GPa. A remarkable property of this phase diagram is its strong sensitivity on the pressure conditions. As shown by the open symbols in Fig. 1, $T_c$ is increased when the pressure is applied on solid helium by comparison with liquid helium. As expected, the effect is even more dramatic with less hydrostatic media. When using a 1:1 mixture of Fluorinert FC70:FC77, $T_c$ is only slowly reduced with pressure and $T_c \approx 3.19$ K at our pressure limit of 5.8 GPa [28]. In our dc magnetization measurements using Daphne 7474, a second superconducting dome is even obtained with a maximum $T_c$ as high as 3.8 K at 5.5 GPa, which is above the room temperature solidification point of this medium [28]. Such a large sensitivity to the hydrostaticity is most likely a consequence of the anisotropic uniaxial pressure dependence of $T_c$. In $KFe_2As_2$, $\partial T_c/\partial p_a|_0 \approx -1.9$ K GPa$^{-1}$ along the $a$ axis, whereas $\partial T_c/\partial p_c|_0 \approx +2.1$ K GPa$^{-1}$ along the $c$ axis [49]. Under hydrostatic conditions, the three axes will contribute equally to give rise to the phase diagram presented in Fig. 1. However, under nonhydrostatic conditions, as already explained in Refs. [32, 42], the pressure will be larger along the $c$ axis and smaller in the $ab$ plane. This results in larger values of $T_c$ and a modification of the superconducting phase diagram [28].

Not only do we confirm the kink in $T_c(p)$ previously reported [9, 16], but we observe that $T_c$ remains roughly constant up to 7.1 GPa. A Lifshitz transition can produce such a kink in $T_c(p)$. In that case, the observed increase of $T_c$ just above $p^*$ is consistent with the formation of a new Fermi-surface pocket [50]. However, no anomaly was observed in the Hall coefficient to support this mechanism [9], and dHvA oscillations indicate no drastic change in the Fermi surface up to $\sim 2.5$ GPa [16].

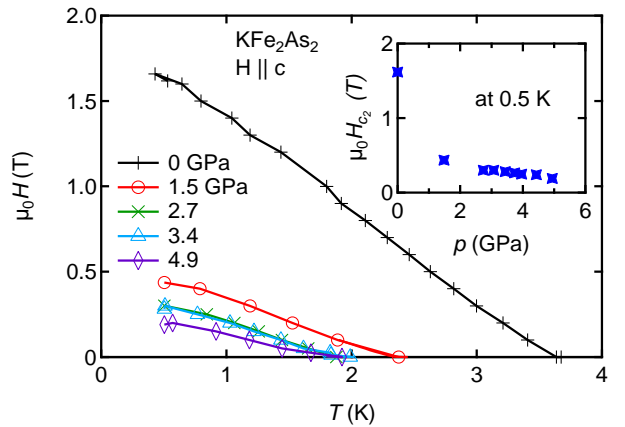

Figure 2. (Color online) Temperature dependence of the upper critical field $H_{c_2}$ at different pressures. The data were determined using a resistive transition offset criteria ($\rho = 0$, see Ref. [25]). We note that, unlike the use of the transition midpoint criteria, the chosen offset criteria agrees more closely with ambient pressure specific heat measurements [51]. The inset shows the pressure dependence of $H_{c_2}$ at 0.5 K.

We note that the change in slope at the characteristic pressure of 1.8 GPa could be a simple consequence of the fact that the uniaxial pressure dependencies of $T_c$ are of opposite sign and large. At ambient pressure, $\partial T_c/\partial p_a|_0 \approx -1.9$ K GPa$^{-1}$ and $\partial T_c/\partial p_c|_0 \approx +2.1$ K GPa$^{-1}$ in Ref. [49] or $\partial T_c/\partial p_c|_0 \approx +1.1$ K GPa$^{-1}$ in Ref. [42]. These partial derivatives cancel each other to a considerable degree, yielding a hydrostatic pressure derivative that is negative. Any nonlinearity in $T_c(p_a)$ or $T_c(p_c)$ would generate a much larger relative nonlinearity in the dependence of $T_c$ on hydrostatic pressure $T_c(p)$. For example, were the magnitude of $\partial T_c/\partial p_a$ to gradually decrease by a factor of $\sim 3$ under 3 GPa hydrostatic pressure, $\partial T_c/\partial p_c$ remaining constant, the hydrostatic pressure dependence $T_c(p)$ would be forced to pass through a minimum. We also note that, even though the modulus of elasticity is almost identical along the $a$ and $c$ axis, the first derivative of the modulus is over an order of magnitude smaller along the $c$ axis [17]. This implies a larger compression along the $c$ axis, so that the effect of pressure on $T_c$ may become dominated by the $p_c$ component. In such a scenario, a theoretical explanation of the uniaxial pressure dependencies of $T_c$ would be the key to understanding the slope change at $p*$.

Another possibility that may even induce the previous idea is a transition to a superconducting phase of a different symmetry. In such a case, changes in other thermodynamic quantities, such as the thermal expansion or the specific heat, are also expected. However, the combination of high pressures and low temperatures makes the experimental investigations of these quantities challenging. Figure 2 shows the temperature dependence of $H_{c_2}$ for the magnetic field applied along the $c$ axis at different pressures. At ambient pressure, the upper critical field along the $c$ axis is known to be due to the orbital limit with negligible effect due to the Pauli limit [49, 52]. With increasing pressure, as $T_c$ decreases, $H_{c_2}$ is also decreasing. Interestingly, above $p*$ where $T_c$ remains roughly constant or increases very slowly, $H_{c_2}$ continues to decrease (see inset of Fig. 2). In the following, we will relate this decrease with a commensurate decrease of the electrons' effective mass.

Figure 3 shows the low-temperature dependence of the resistivity as a function of $T^2$ at various pressures (full lines). For each pressure, we performed fits with a Fermi liquid behavior $\rho = \rho_0 + AT^2$ up to 8 K (dashed lines). The pressure dependence of the $A$ coefficient is shown in the inset. At ambient pressure, $A \approx 0.02\ \mu\Omega$ cm $K^{-2}$, in agreement with previous reports [5, 52–54]. Under pressure, $A$ decreases smoothly, which is consistent with the decreasing trend in effective mass observed in dHvA oscillations [16].

In Ref. 55, the Helfand-Werthamer theory is examined for the case of uniaxial anisotropy with an anisotropic superconducting gap. For $H||c$ (one-band case, clean limit):

$$\frac{1}{T_c}\left(-\left.\frac{d\mu_0 H_{c_2}^{\mathrm{orb}}}{dT}\right|_{T_c}\right) = \frac{8}{7\zeta(3)\langle\Omega^2\mu_c\rangle}\frac{\phi_0 2\pi k_B^2}{\hbar^2 v_0^2} \quad (1)$$

The function $\Omega(\mathbf{k_F})$ which determines the $\mathbf{k_F}$ dependence of the superconducting gap $\Delta = \Psi(\mathbf{r},T)\Omega(\mathbf{k_F})$ is normalized so that $\langle\Omega^2\rangle = 1$. The averages over the Fermi surface are shown as $\langle\cdots\rangle$.

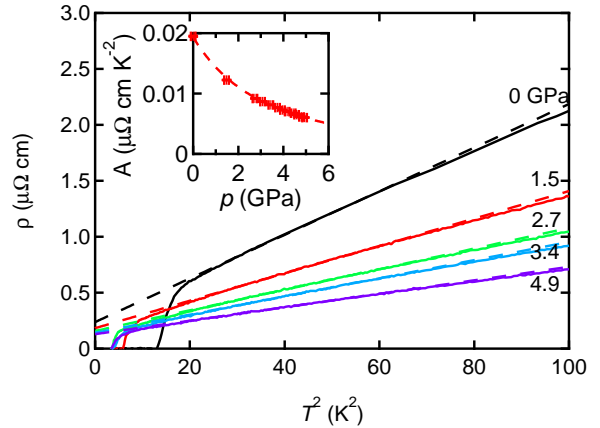

Figure 3. (Color online) Electrical resistivity $\rho$ of KFe$_2$As$_2$ at different pressures at low temperatures as a function of $T^2$. The dashed lines show fits of the resistivity to a Fermi liquid behavior $\rho = \rho_0 + AT^2$. The pressure dependence of the $A$ coefficient is shown in the inset. The dashed line represents $A(0)/(1+\beta p)^2$ where $\beta = 0.16(1)$ GPa$^{-1}$.

$$\mu_c = \frac{v_x^2 + v_y^2}{v_0^2} \quad \text{and} \quad v_0^3 = \frac{2E_F^2}{\pi^2\hbar^3 N(0)}$$

where $N(0)$ is the total DOS at the Fermi level $E_F$ per spin. Assuming that the $A$ coefficient of the $T^2$ term of the resistivity, when resistivity is measured along $x$, obeys $A \propto n/\langle v_x^2\rangle$, the $A$ coefficient in the tetragonal case will reflect the dependence on $1/(v_0^2\langle\mu_c\rangle)$:

$$\frac{1}{T_c}\left(-\left.\frac{d\mu_0 H_{c_2}^{\mathrm{orb}}}{dT}\right|_{T_c}\right) \propto \frac{\langle\mu_c\rangle}{\langle\Omega^2\mu_c\rangle}\frac{A}{n} \quad (2)$$

where the carrier density $n$ can be estimated from Hall measurements. A more detailed expression for $A$ can be found in various publications [56, 57] and would lead to a more complicated expression than that given in Eq. 2. A refinement to the case of several bands would certainly be of interest. In the present form, Eq. 2 shows a proportionality between the slope of $H_{c_2}$ at $T_c$ and the $A$ coefficient. This result is known for heavy fermions, both quantities being proportional to the square of the effective mass [58, 59].

Figure 4 shows the plot of $\left(-d\mu_0 H_{c_2}/dT|_{T_c}\right)/T_c$ versus $A$. All measured points fall onto two straight lines of different slope corresponding to $p < p*$ and $p > p*$. It is remarkable that both lines go through the origin as expected from the proportionality relation in Eq. 2. Equation 2 indicates that a change of slope when plotting $\left(-d\mu_0 H_{c_2}^{\mathrm{orb}}/dT|_{T_c}\right)/T_c$ versus $A$ implies a change in either $n$, $\Omega$, or $\mu_c$. In KFe$_2$As$_2$, the carrier density does not change significantly with pressure as inferred from Hall resistivity measurements [9] and from the smooth pressure variation of the $A$ coefficient. Therefore, the observed change of slope is more likely due to a change in $\langle\Omega^2\mu_c\rangle$. We note that a change of the superconducting gap symmetry is not the only possible explanation, since a change of the Fermi surface will also modify the value of $\langle\Omega^2\mu_c\rangle$. However, dHvA oscillations experiments indicate that the global structure of the Fermi surface hardly changes up to $p \sim 2.5$ GPa. We mention that $\langle\mu_c\rangle/\langle\Omega^2\mu_c\rangle = 1$ for any $\Omega$ that does not depend on $k_z$. Therefore, a simple change between $s$ wave and $d$ wave would not be able to explain the change of slope of nearly a factor of two observed in Fig. 4.

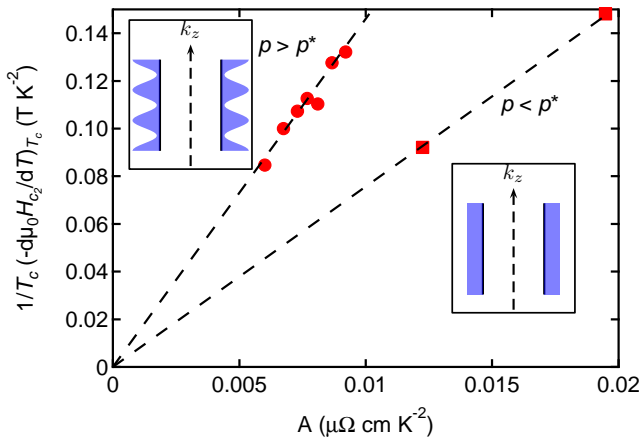

Figure 4. (Color online) Plot of $\left(-d\mu_0 H_{c_2}/dT|_{T_c}\right)/T_c$ vs the $A$ coefficient of resistivity. Each point corresponds to a different pressure. All measured points fall onto two lines of different slope corresponding to $p < p*$ and $p > p*$. The two insets show a schematic gap structure with the appearance of a modulation along $k_z$ above $p^*$.

On the other hand, the appearance of a $k_z$ modulation of the superconducting gap at $p^*$ can explain an increase of nearly a factor two in $\langle\mu_c\rangle/\langle\Omega^2\mu_c\rangle$. Let us assume a superconducting gap with a modulation along $k_z$ [55, 60, 61]:

$$\Delta = \Delta_0\left(1 + \eta\cos\left(k_z c^*\right)\right), \quad (3)$$

where $c^* = \hbar v_0/(2E_F)$ is the length scale [55]. Let us also assume a prolate ellipsoidal Fermi surface ($\epsilon = 0.1$ in the notations of Ref. 55). We find that $\langle\mu_c\rangle/\langle\Omega^2\mu_c\rangle$ changes from 1 to $\sim 1.9$ if $\eta$ changes from 0 to $-0.8$. Therefore, if we assume that the Fermi surface does not change at $p^*$, the appearance of a $k_z$ modulation of the superconducting gap at $p^*$ can be a possible explanation of our experimental observations. A $k_z$ dependence of the superconducting gap has been observed by ARPES in $Ba_{0.6}K_{0.4}Fe_2As_2$ [61], in agreement with a theoretical prediction for the pairing strength [62]. The pairing between the layers is predominantly responsible for the gap dispersion with $k_z$. In contrast, in $Ba_{0.1}K_{0.9}Fe_2As_2$, the superconducting gap size on all the $\Gamma$-centered hole Fermi surfaces does not vary much along $k_z$ [63]. This is consistent with the near two-dimensionality of $KFe_2As_2$ by comparison with the other members of the 122 family [22, 64–67]. It is possible that, by applying pressure on $KFe_2As_2$, the pairing between the layers induces a $k_z$ modulation of the superconducting gap. This is also consistent with dHvA oscillations measurements showing that three-dimensionality increases with pressure [16].

In conclusion we have shown that there is very likely a change in the $k_z$ modulation of the SC gap at $p^* \sim 1.8$ GPa. We base this conclusion on a change in the scaling of $\left(-d\mu_0 H_{c_2}/dT|_{T_c}\right)/T_c$ with the $A$ coefficient of the $T^2$ term of the resistivity. We have shown that this indicates either a change of the Fermi surface, of the carrier density, and/or of the superconducting gap symmetry. In addition, we significantly extended the pressure temperature phase diagram from $\sim 2.5$ to $7.1$ GPa. For $p > 2.5$ GPa we found that $T_c$ increases only slightly up to $7.1$ GPa. By using various pressure cells and several different pressure-transmitting media, we have demonstrated the extreme sensitivity of $KFe_2As_2$ to nonhydrostaticity [28] and propose that it is due to the anisotropic dependence of $T_c$ on strain.

We would like to thank A. Jesche, L. Howald, D. Finnemore, T. Kong, and F. F. Tafti for useful discussions. This work was carried out at the Iowa State University and supported by AFOSR-MURI Grant No. FA9550-09-1-0603. Part of this work was performed at the Ames Laboratory, US DOE, under Contract No. DE-AC02-07CH11358. The ac susceptibility measurements in a diamond anvil cell were carried out at Washington University in St. Louis and supported by the National Science Foundation (NSF) through Grant No. DMR-1104742 and by the Carnegie/DOE Alliance Center (CDAC) through NNSA/DOE Grant No. DE-FC52-08NA28554.

# Supplementary Material for "Upper critical field of $KFe_2As_2$ under pressure: a test for the change in the superconducting gap structure"

Valentin Taufour,[1, 2, *] Neda Foroozani,[3] Makariy A. Tanatar,[1, 2] Jinhyuk Lim,[3] Udhara Kaluarachchi,[1] Stella K. Kim,[1, 2] Yong Liu,[2] Thomas A. Lograsso,[2] Vladimir G. Kogan,[2] Ruslan Prozorov,[1, 2] Sergey L. Bud'ko,[1, 2] James S. Schilling,[3] and Paul C. Canfield[1, 2]

[1] *Department of Physics and Astronomy, Iowa State University, Ames, Iowa 50011, U.S.A.*
[2] *The Ames Laboratory, US Department of Energy, Iowa State University, Ames, Iowa 50011, USA*
[3] *Physics Department, Washington University, St. Louis, Missouri 63130, U.S.A.*

## SUPPLEMENTARY EXPERIMENTAL DETAILS

For the electrical resistivity measurements under pressure, we used a 1:1 mixture of n-pentane:isopentane as a pressure medium. The solidification of this medium occurs at room temperature at $\sim 6-7$ GPa. [1, 2] Another experiment was performed with a 1:1 mixture of Fluorinert FC70:FC77 which solidifies at room temperature at $\sim 0.8 - 1$ GPa [3, 4]. The pressure was determined from the value of the superconducting transition of Pb in the electrical resistivity measured by the four-probe method [5]. The electrical resistivity of $KFe_2As_2$ was measured by the four-probe method with current in the ab-plane. The realization of four electrical contacts on small samples of $KFe_2As_2$ is difficult. Attempts with spot welding, silver paint or silver epoxy have resulted in electrical contacts too weak to survive the rigors of a pressure loading. Four silver wires (25 $\mu$m diameter) were finally soldered on the sample.[6] Measurements down to 2 K were performed in a commercial Quantum Design Physical Property Measurement System (PPMS), and measurements down to 0.4 K were performed in a $^3$He cryostat (CRYO Industries of America) with a Linear Research LR700 ac resistance bridge. The temperature was measured with a Cernox thermometer positioned on the body of the pressure cell at the level of the sample position. Due to non negligible contact dimensions compared to the small sample size, the absolute resistivity was measured on several bigger samples at ambient pressure and a value of $(300 \pm 30)$ $\mu\Omega \cdot$cm was obtained. All pressure dependent data were normalized to this ambient pressure, room temperature value.

For the ac susceptibility measurements, the diamond anvil cell has two opposing 1/6-carat, type-Ia diamonds with either 0.5 mm or 0.9 mm diameter culets. The primary ac field was 3 Oe (rms) at 1023 Hz. The signal from two calibrated, compensated secondary coils was fed into an SR554 transformer preamplifier connected to a Stanford Research SR830 digital lock-in amplifier. Temperatures were measured using a calibrated Cernox resistor (Lake Shore Cryotronics) located close to the sample. A 3 mm diameter gold-plated CuBe gasket was preindented and the cylindrical sample space (half the culet diameter) spark-cut through the center. Small ruby spheres [7] were placed in the sample space to determine the given values of pressure in situ at 5-10 K with a resolution of $\pm 0.1$ GPa using the revised pressure scale of Chijioke et al. [8] Further details of the pressure techniques are given elsewhere [9, 10].

For the magnetization measurement under pressure, we used a moissanite anvil cell [11]. The body of the cell is made of Cu-Ti alloy and the gasket is made of Cu-Be. The pressure was applied at room temperature and Daphne 7474 was used as a pressure transmitting medium. The solidification of this medium occurs at room temperature at $\sim 3.7$ GPa [12]. The pressure was determined at 77 K by the ruby fluorescence technique [13]. The temperature dependent, zero field-cooled (ZFC) magnetization was measured under pressure in a Quantum Design MPMS-SQUID magnetometer down to 1.9 K in a magnetic field of 15 Oe, 50 Oe and 100 Oe applied along the $c$-axis.

## SUPPLEMENTARY MATERIAL ON THE $T$-$p$ PHASE DIAGRAM

### Electrical resistivity

The superconducting phase transition at low temperatures was studied via electrical resistivity, magnetization and ac susceptibility measurements. Figure 1(a) shows the temperature dependence of the electrical resistivity at different pressures with a 1:1 mixture of $n$-pentane:isopentane as a pressure medium. A complete superconducting transition is observed for all measured pressures and we define the superconducting transition temperature $T_c$ when zero resistivity is obtained. At ambient pressure, $T_c = 3.61$ K and $T_c$ initially decreases with applied pressure. At 2.7 GPa, $T_c$ reaches 1.95 K and remains roughly constant up to our pressure limit of 4.9 GPa. To gain insight on the influence of non-hydrostaticity of the pressure medium on the phase diagram of $KFe_2As_2$, we performed another resistivity experiment using a 1:1 mixture of Fluorinert FC70:FC77 as a pressure medium. Figure 1(b) shows the temperature dependence of the electrical resistivity at different pressures with that pressure medium which is a much less hydrostatic medium. The superconducting transition of $KFe_2As_2$ is complete for all measured pressures. $T_c$ is much more slowly reduced with the application of pressure and, at our pressure limit of 5.8 GPa, $T_c = 3.19$ K.

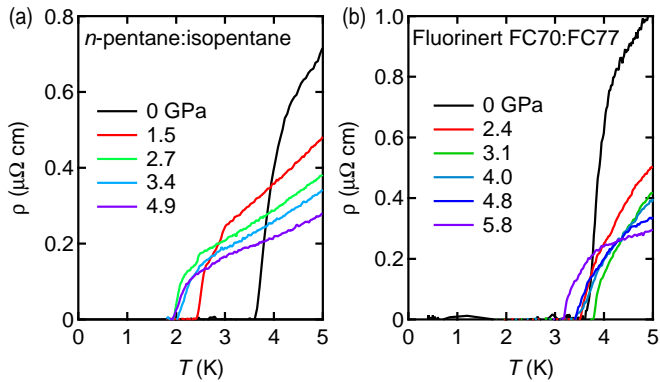

Figure 1. (Color online) Temperature dependence of the electrical resistivity of $KFe_2As_2$ at different pressures at low temperatures, measured in a modified Bridgman cell [14] with two different pressure media: (a) a 1:1 mixture of $n$-pentane:isopentane and (b) a 1:1 mixture of Fluorinert FC70:FC77.

## Magnetization

We also studied the superconducting properties of $KFe_2As_2$ under pressure by low field magnetization measurements. Figure 2 shows the temperature dependence of the ZFC magnetization at different pressures for two different samples. At 15 Oe, both sample #1 and #2 show the superconducting transition with an onset at $T_c = 3.49(2)$ K. Upon applying pressure, $T_c$ initially decreases. At 2.7 GPa, we only see the onset of the superconducting transition because our measurements are limited to temperatures above 1.9 K. The most striking result is that the superconducting transition temperature increases up to 3.8 K at 5.5 GPa. A further increase of the pressure to 6.3 GPa leads to a sharper transition. At 6.3 GPa, the shielding volume fraction at 2 K is $\sim$ 83% of its value at ambient pressure. This indicates that more than 80% of the surface of the sample is superconducting. In the case of filamentary superconductivity, a zero resistance value can be obtained but no shielding fraction is visible in the magnetization, as can be seen for example in some $Ca(Fe_{1-x}Co_x)_2As_2$ samples [15]. At higher pressure, the superconducting transition temperature is suppressed down to 2.46 K at our pressure limit of 6.8 GPa.

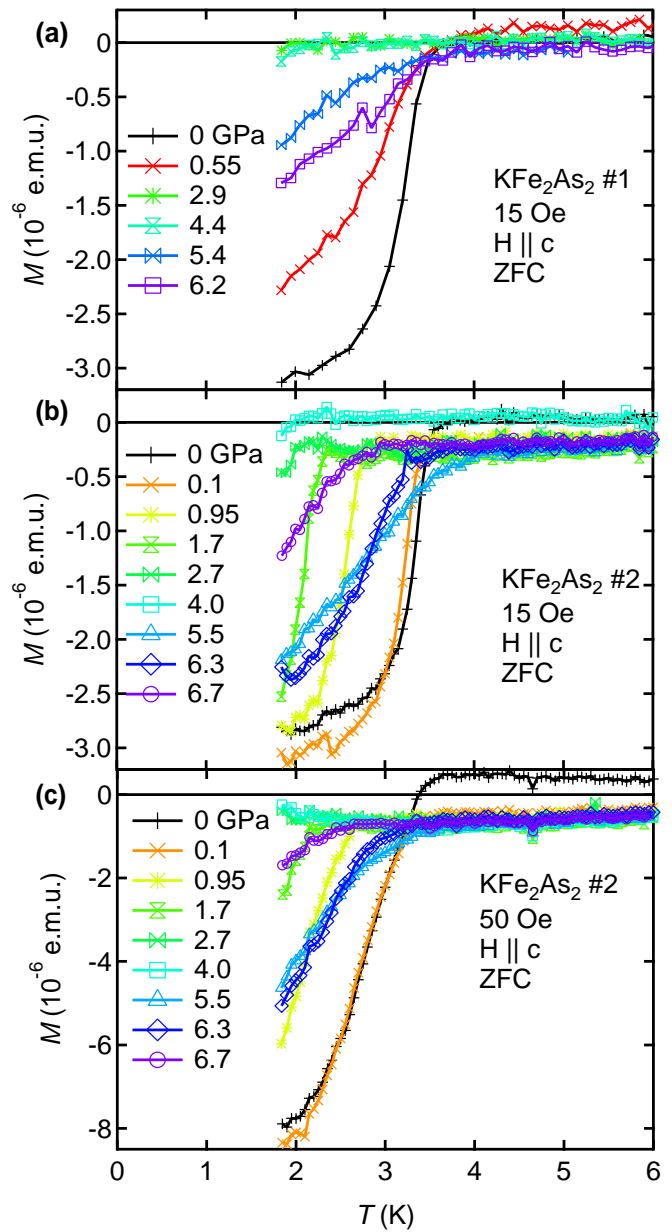

Figure 2. (Color online) Temperature dependence of the ZFC magnetization at different pressure measured at 15 Oe on sample#1 (a) and sample#2 (b) and at 50 Oe on sample#2 (c) with the magnetic field along the $c$-axis.

## AC Susceptibility

We performed three separate ac susceptibility measurements (runs #1,2,3) on $KFe_2As_2$ in a diamond anvil cell with helium as pressure transmitting medium. In Fig. 3(a) we show the superconducting transition at ambient pressure as revealed by the in-phase part of the 1st harmonic $\chi'_1$, in Fig. 3(b) the out-of-phase part of the 1st harmonic $\chi''_1$, and in Fig. 3(c) the out-of-phase part of the 3rd harmonic $\chi''_3$. These measurements are taken in the pressure cell without applied load. It is seen that the temperature of the midpoint of the superconducting transition in $\chi'_1$ lies very close to the temperature of the maximum in either $\chi''_1$ or $\chi''_3$. Since below 4 K the temperature dependence of the background signal in $\chi''_1(T)$ or $\chi''_3(T)$ is far weaker than that for $\chi'_1(T)$, a more accurate estimate of the value of the superconducting transition temperature $T_c$ can be obtained from $\chi''_1(T)$ or $\chi''_3(T)$.

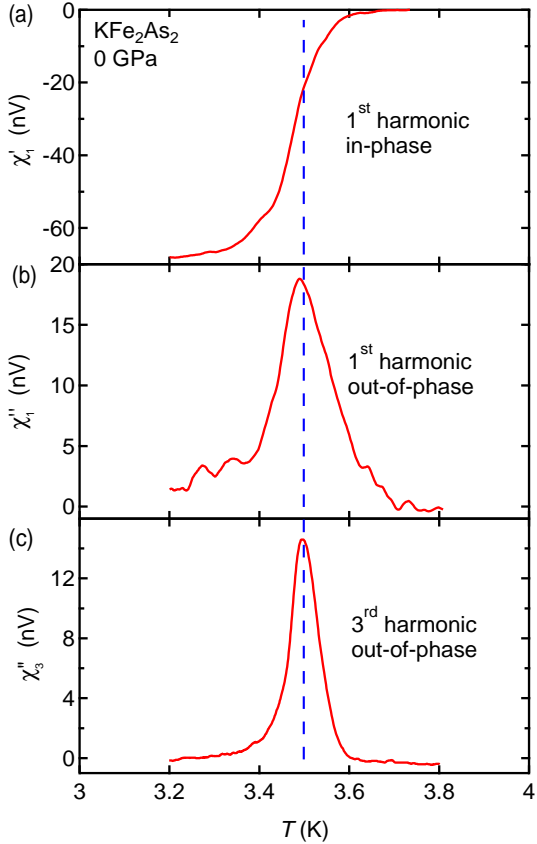

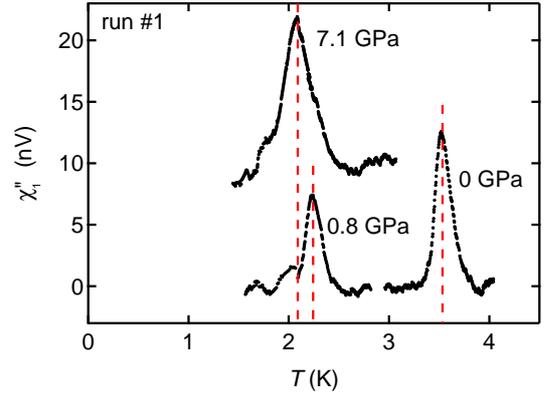

Figure 3. (Color online) Superconducting transition in ac susceptibility of $KFe_2As_2$ at ambient pressure: (a) in-phase part of 1st harmonic, (b) out-of-phase part of 1st harmonic, (c) out-of-phase part of 3rd harmonic.

In Fig. 4, $\chi''_1$ from run #1 is plotted versus temperature at 0, 0.8, and 7.1 GPa. In this run diamond anvils with 0.5 mm diameter culets were used. To facilitate obtaining pressures in the pressure range below 7.1 GPa, anvils with the larger culet diameter of 0.9 mm were chosen for runs #2 and 3. $\chi''_1(T)$ was used to determine $T_c(p)$ in run #2, whereas in run #3 $\chi''_3(T)$ was measured, as shown in Fig. 5. This figure shows clearly how $T_c$ initially decreases with applied pressure, reaches a minimum of $\sim 1.83$ K at 1.78 GPa, and increases only slowly at higher pressures. The change of pressure from 4.32 GPa to 5.77 GPa was made at 10 K where the helium pressure medium was solid. This resulted in a broader transition due to a larger pressure gradient and in a slightly higher $T_c$ of 2.23 K. However, this is in sharp contrast with the $T_c$ of 3.8 K at 5.5 GPa in the magnetization measurements in which Daphne 7474 was used as pressure medium.

Figure 4. (Color online) Superconducting transition in out-of-phase part of ac susceptibility (1st harmonic $\chi''_1$) at several pressures in run #1. Data are shifted vertically for clarity.

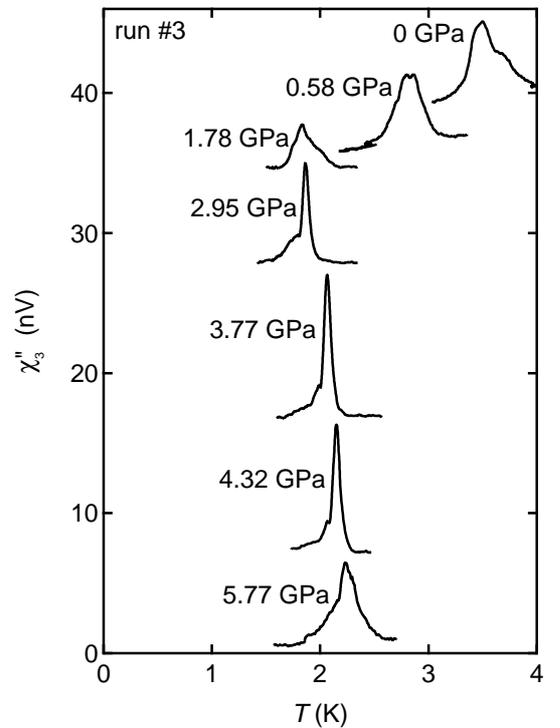

Figure 5. (Color online) Superconducting transition in out-of-phase part of ac susceptibility (3rd harmonic $\chi''_3$) at several pressures in run #3. Data are shifted vertically for clarity.

## Comparison of the $T$-$p$ phase diagram from different experiments: effect of the pressure medium

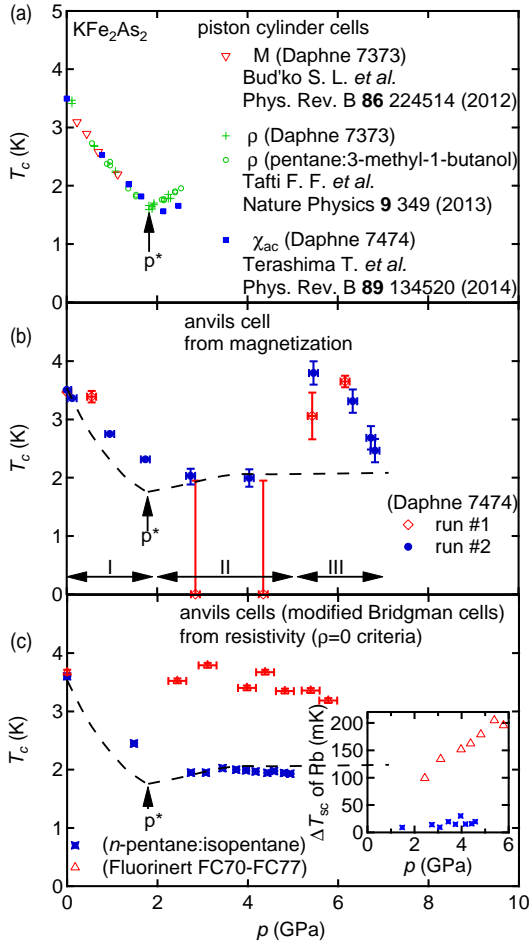

Figure 6. (Color online) Superconducting phase diagram of $KFe_2As_2$ determined from different experiments: (a) using piston cylinder pressure cells from magnetization (data from Ref. 16), from resistivity in different pressure media (data from Ref. 17) and from ac susceptibility (data from Ref. 18); (b) using anvil cell from magnetization on two different samples; (c) using anvil cells (modified Bridgman cells [14]) from resistivity in different pressure media (the inset shows the width of the superconducting transition of Pb in the different pressure media); The dashed line shows the phase diagram determined in the best pressure conditions (average of the data from piston cylinder cells with liquid medium below 2 GPa and data from diamond cell with liquid helium as a pressure medium above 2 GPa, see main article) and serves for comparison between the different phase diagrams.

The previously published temperature versus pressure phase diagrams are reproduced in Fig. 6(a) for comparison with our results. In all Refs. 16–18, a piston cylinder cell was used and the agreement between their results is good at least up to 1.8 GPa. At $p^* \approx 1.8$ GPa, a change of slope in $T_c(p)$ is observed. It was remarked in Ref. 18 that the slope change is rather weak and gradual. The diagram from our magnetization measurements is presented in Fig. 6(b). Three pressure ranges (I, II and III) can be discerned. In region I ($p < p^*$), $T_c$ decreases from 3.5 to 2 K. In region II ($p^* - 4.5$ GPa), $T_c \approx 2$ K and remains roughly constant. In region III ($5.4 - 6.8$ GPa) a new superconducting dome is observed with a maximum $T_c$ of 3.8 K at 5.5 GPa. Figure 6(c) shows the diagram from our resistivity measurements in two different pressure media. When using a 1:1 mixture of $n$-pentane:isopentane, the agreement with our magnetization data is very good for regions I and II (resistivity data did not extend to region III). However, when using a 1:1 mixture of Fluorinert FC70:FC77, $T_c$ is only slowly reduced with pressure. This drastic difference between the two phase diagrams is most likely a consequence of the pressure gradients which are smaller in more hydrostatic media. In our resistivity measurements, the pressure gradient can be estimated from the width of the superconducting transition ($\Delta T_c$) of Pb which is used to determine the pressure at low temperatures. $\Delta T_c$ of Pb is shown in the inset of Fig. 6(c) for the two different pressure media. It is clear that the 1:1 mixture of $n$-pentane:isopentane provides more hydrostatic pressure conditions compared to the 1:1 mixture of Fluorinert FC70:FC77. Figure 6(c) demonstrates the importance of the pressure medium in studying the phase diagram of $KFe_2As_2$. In Figs. 6(b) and (c), the dashed line shows the phase diagram determined in the best pressure conditions (average of the data from piston cylinder cells with liquid medium below 2 GPa and data from diamond cell with liquid helium as a pressure medium above 2 GPa, see the phase diagram in the main article) and serves for comparison between the different phase diagrams.

In this Supplementary Material, we will discuss two other observations from the phase diagrams in Fig. 6. First, in the pressure range $0 - 1.8$ GPa (region I), the decrease of $T_c$ is slightly different between various experiments. Second, we observed a superconducting dome in the range $5.4 - 6.8$ GPa (region III) in our magnetization measurement using Daphne 7474 as a pressure medium.

The first point can be understood when noticing essentially two values for $dT_c/dp$. When piston cylinder cells are used [16–18], or in our study using a diamond anvil cell with helium as a pressure medium, $dT_c/dp \approx -1.1(1)$ K GPa$^{-1}$. By contrast, $dT_c/dp \approx -0.7(1)$ K GPa$^{-1}$ when anvil-based pressure cells are used (moissanite anvil cell and modified Bridgman cell in this study). This difference can be understood considering that a large pressure gradient exists in the direction parallel to the load [4]. As already mentioned in Ref. 16, anvil-based pressure cells are more likely to generate uniaxial pressure gradients compared to piston cylinder pressure cells. Due to the plate-like shape of the $KFe_2As_2$ samples, the $c$-axis is perpendicular to the anvil culet. Consequently, the pressure along the $c$-axis in such a cell is larger than the pressure in the *ab* plane. The fact that the phase diagram can be different depending on the orientation of the sample inside the pressure cell has already been demonstrated [19]. As we can see from our ac susceptibility measurements in a diamond anvil cell with helium as a pressure medium, the uniaxial pressure gradient in anvil-based pressure cells can be significantly reduced by using a highly hydrostatic medium such as liquid helium. Daphne 7474 or *n*-pentane:isopentane are usually considered as hydrostatic media below 3 GPa, but our study shows that these media are significantly less hydrostatic than helium in anvil-based pressure cells. Due to the much larger sample space in piston-cylinder cells, the uniaxial pressure gradient is smaller. However, we note that the effect of the uniaxial pressure gradient can also be observed in those pressure cells, when the sample properties are anisotropic [20]. Therefore, the slope $dT_c/dp$ is more accurately obtained in piston cylinder cells (Refs. 16–18) or in anvil cells with helium as a pressure medium (this study) than in anvil cells with less hydrostatic liquid media (Daphne 7474 or *n*-pentane:isopentane) due to a larger uniaxial component in anvil-based pressure cells. Using the same argument, we understand that the kink in $T_c(p)$ at 1.8 GPa is shifted to $\sim 2$ GPa in our study with Daphne 7474 or *n*-pentane:isopentane in anvil-based pressure cells. The advantage of anvil-based pressure cells remains that larger pressures can be obtained.

The apparent high-pressure superconducting dome (region III) is most likely an artifact associated with non hydrostaticity and serves again to underscore the profound sensitivity of these materials to details of the pressure environment [4, 16, 19–24]. Although we observe the reappearance of superconductivity with a clear signal in magnetization in two different samples, we cannot confirm the existence of a new intrinsically superconducting state by resistivity because we were not able to reach the same pressure range in our resistivity measurements. In addition, this new dome was not observed in the ac susceptibility measurements with helium as a pressure medium. It is important to recall, though, that the solidification of Daphne 7474 occurs at room temperature at $\sim 3.7$ GPa [12]. As such then, a large uniaxial pressure gradient is most likely responsible for the observed superconductivity. This is supported by the fact that $T_c$ does not decrease much when using a 1:1 mixture of Fluorinert FC70:FC77, which solidifies at room temperature at $\sim 0.8 - 1$ GPa [3, 4]. In fact, the values of $T_c$ for the non-hydrostatic data in resistivity (Fig. 6(c)) are comparable to the maximum of the dome in Fig. 6(b). As explained in the main text, all the higher $T_c$ values obtained under less hydrostatic conditions are consistent with the anisotropic uniaxial pressure dependencies of KFe$_2$As$_2$ [16, 25].

## SUPPLEMENTARY TEMPERATURE DEPENDENCE OF THE ELECTRICAL RESISTIVITY

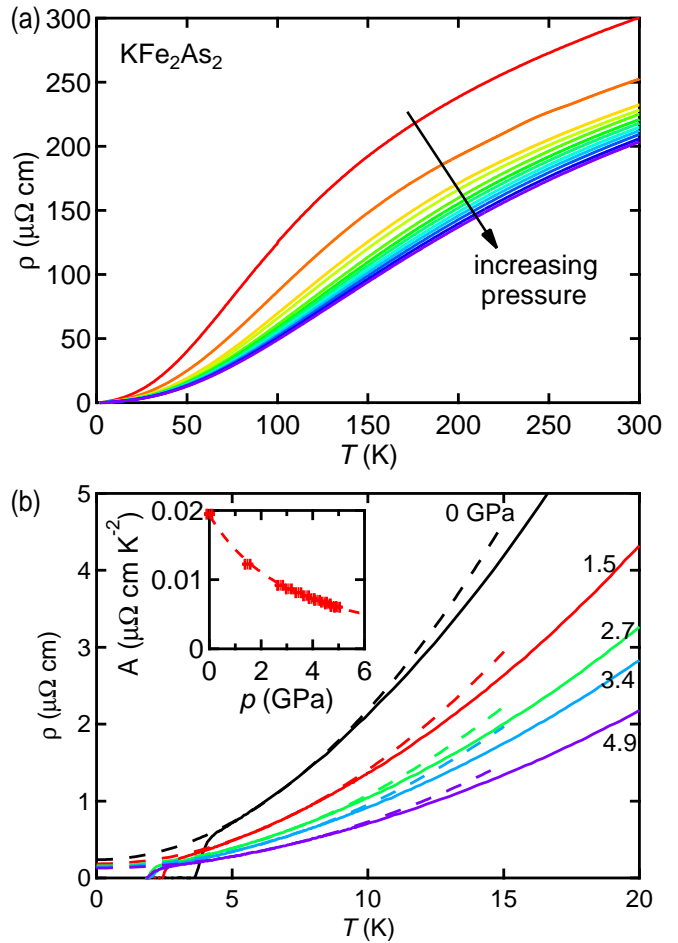

Figure 7. (Color online)(a) Temperature dependence of the electrical resistivity $\rho$ of KFe$_2$As$_2$ at different pressures up to 4.9 GPa. The pressure medium is a 1:1 mixture of *n*-pentane:isopentane. (b) Electrical resistivity $\rho$ of KFe$_2$As$_2$ at different pressures at low temperatures as a function of temperature. The dashed lines show fits of the resistivity to a Fermi liquid behavior $\rho = \rho_0 + AT^2$. The pressure dependence of the $A$ coefficient is shown in the inset. The dashed line represents $A(0)/(1 + \beta p)^2$ where $\beta = 0.16(1)$ GPa$^{-1}$.

Figure 7(a) shows the temperature dependence of the electrical resistivity at different pressures. At ambient pressure, our results confirm the previous reports [26–29] on the resistivity of KFe$_2$As$_2$ in which the Fermi liquid $T^2$ dependence has been observed at low temperatures. The residual resistivity ratio (RRR) of our sample is $\sim 1200$, similar to the best values reported so far. Figure 7(b) shows the low temperature dependence of the resistivity as a function of temperature at various pressures (full lines). For each pressure, we performed fits to a Fermi liquid behavior $\rho = \rho_0 + AT^2$ up to 8 K (dashed lines).

We found that a deviation from a $T^2$ behavior starts above $\sim 9$ K, which is lower than $\sim 45$ K reported in Ref. 26. Since the electronic correlations (and therefore the $A$ coefficient) are rather large, the temperature $T_{Fl}$, below which the $T^2$ behavior can be observed, is expected to be low. In addition, the rather low Debye temperature implies that the phonon contribution is expected to be significant down to low temperatures and further decrease $T_{Fl}$. Our data indicate that the $T^2$ behavior is a good fit below 8 K although we cannot exclude other exotic behavior (see Refs. 30–32 for a discussion on the $T^n$ behavior where sample dependencies are also observed). Here we focus on the pressure dependence of the $A$ coefficient shown in the inset of Fig. 7(b). At ambient pressure, $A \approx 0.02$ $\mu\Omega$ cm $K^{-2}$, in agreement with previous reports [26, 28]. Under pressure, $A$ decreases smoothly, which is consistent with the decreasing trend in effective mass observed in dHvA oscillations [18].